\documentclass[preprint,12pt]{elsarticle}

\usepackage{amssymb}

\usepackage{booktabs}
\usepackage{multirow}
\usepackage{subcaption}
\usepackage{graphicx}
\usepackage[hyphens]{url}
\usepackage{lscape}
\usepackage{pifont}
\usepackage{amsmath}

\journal{Computers \& Security}

  \makeatletter
  \def\ps@pprintTitle{%
    \let\@oddhead\@empty
    \let\@evenhead\@empty
    \def\@oddfoot{\reset@font\hfil\thepage\hfil}
    \let\@evenfoot\@oddfoot
  }
  \makeatother

\begin{document}

\begin{frontmatter}



\title{Generalizing intrusion detection for heterogeneous networks: A stacked-unsupervised federated learning approach}

\author[1]{Gustavo de Carvalho Bertoli\corref{cor1}}
\ead{bertoli@ita.br}

\author[1]{Lourenço Alves Pereira Junior}
\ead{ljr@ita.br}

\author[1]{Osamu Saotome}
\ead{osaotome@ita.br}

\author[2]{Aldri Luiz dos Santos}
\ead{aldri@dcc.ufmg.br}

\cortext[cor1]{Corresponding author}

\affiliation[1]{
    organization={Aeronautics Institute of Technology (ITA)},
    addressline={Praca Marechal Eduardo Gomes, 50}, 
    city={S\~{a}o Jos\'{e} dos Campos},
    postcode={12228-900}, 
    state={SP},
    country={Brazil}
}
    
\affiliation[2]{
    organization={Federal University of Minas Gerais (UFMG)},
    addressline={Av. Antônio Carlos, 6627, Pampulha}, 
    city={Belo Horizonte},
    postcode={31270-901}, 
    state={MG},
    country={Brazil}
}

\begin{abstract}
    The constantly evolving digital transformation imposes new requirements on our society.  Aspects relating to reliance on the networking domain and the difficulty of achieving security by design pose a challenge today.   As a result, data-centric and machine-learning approaches arose as feasible solutions for securing large networks.
    Although, in the network security domain, ML-based solutions face a challenge regarding the capability to generalize between different contexts.  In other words, solutions based on specific network data usually do not perform satisfactorily on other networks.
    This paper describes the stacked-unsupervised federated learning (FL) approach to generalize on a cross-silo configuration for a flow-based network intrusion detection system (NIDS).
    The proposed approach we have examined comprises a deep autoencoder in conjunction with an energy flow classifier in an ensemble learning task.  
    Our approach performs better than traditional local learning and naive cross-evaluation (training in one context and testing on another network data).  Remarkably, the proposed approach demonstrates a sound performance in the case of \textit{non-IID} data silos.
    In conjunction with an informative feature in an ensemble architecture for unsupervised learning, we advise that the proposed FL-based NIDS results in a feasible approach for generalization between heterogeneous networks. 
\end{abstract}



\begin{keyword}
Network Intrusion Detection \sep Generalization \sep Unsupervised Learning \sep Federated Learning \sep Network flows


\end{keyword}

\end{frontmatter}


\section{Introduction}

The current and envisioned network communications infrastructure is expected to evolve continuously with technological advancements and the introduction of breakthrough services. For example, by 2030, it is predicted that the number of Internet of Things (IoT) devices will come to 500 billion, and 90\% of vehicles will be associated with the IoT~\citep{iot_forecast_2030}. These technological advancements are enablers of IoT verticals such as smart homes, cities, industries, healthcare, and transportation. Although, network security is an emergent property that must be in place by design for trustworthy solutions that can be pervasive in our society~\citep{anderson2020security}.

These networked systems generate a massive amount of data (i.e., network traffic) that presents a prolific path for machine learning (ML) solutions that aim to learn from data in contrast with the traditional algorithmic approach~\citep{abu2012learning}. ML-based Network Intrusion Detection Systems (NIDS) are a mature research domain. However, with advancements in machine learning and emerging techniques such as deep learning and federated learning, new paths in this domain are being explored. 
%
%
A challenge for ML-based NIDS development is pointed out to privacy concerns regarding the network traffic itself~\citep{markusring
}. Better ML-based solutions require data availability, but the sharing of network traffic presents a concern associated with the leak of trade secrets or operational behavior that competitors or malicious agents can misuse. Thus, without the availability and sharing of a diversity of traffics, this data scarcity slows down the advancements of ML-based NIDS. Another challenge for ML-based NIDS is bridging the gap between academic findings and the real operational environment, with most operational solutions still relying on rule-based systems (e.g., Snort). This gap is 
associated with the challenges of generalizing between different networks or addressing network changes (i.e., concept drift).

Federated Learning (FL) is a technique that aims to enable distributed agents to collaboratively learn a certain task without data sharing~\citep{pmlr-v54-mcmahan17a}. The application of FL is a candidate for cyberattack detection in mobile networks~\citep{fl_mobile_survey2020} and applies to edge computing, autonomous driving, and coexistence of heterogeneous systems, among others~\citep{fl_wireless_2020}. Additionally, with the increase in processing capabilities of edge devices, it is envisioned a paradigm shift from the cloud to the crowd with edge devices (the crowd) performing local computations and sharing minimal data (e.g., local model parameters) to a central cloud server as proposed by FL~\citep{ioft}.
%
%
Most intrusion detection research using federated learning (FL) focus on the privacy and efficiency aspects and not on the generalization capabilities achievable with FL. In addition, unsupervised methods for NIDS typically address aspects of unknown attack detection. Our work proposes a setup using FL and unsupervised learning to achieve generalization for NIDS. Despite pointing out the advantage of unsupervised methods over supervised, current works aiming at generalization do not report sound performance to generalize to other contexts (datasets). 

This paper presents an unsupervised FL-based NIDS that can generalize the intrusion detection task to diverse networked systems (cross-silos federated learning). Via a stacking learning setup with a state-of-art anomaly detection algorithm in conjunction with a deep autoencoder. We evaluate our proposal on four recent flow-based NIDS datasets representing different network contexts. The experimental results and empirical analysis demonstrate that our proposed scheme can take advantage of the shared knowledge of its participants through their joint forces for a unified and generalized NIDS solution. Additionally, this FL approach inherits the privacy-preserving decentralized learning aspects from FL.  

Our proposed stacked-unsupervised federated learning approach achieves the best generalization performance between different networks context when compared to state-of-art algorithms and classic unsupervised learning methods. Our method successfully applies unsupervised FL for the problem of network intrusion detection generalization using flow-based data, pushing the research area for generalization on ML-based NIDS.


This paper is organized as follows: In section~\ref{background}, we present a technical background to support understanding network intrusion detection in a federated learning setting. Then, related works are discussed in Section~\ref{related}, presenting both the generalization problem and related FL-based NIDS research. Finally, we describe our methodology in Section~\ref{methodology} with the results and discussion in Section~\ref{results}. 
Finally, we bring conclusions and propose future directions and challenges in Section~\ref{conclusion}.

\section{FL and ML NIDS Background}\label{background}
This section presents the Machine Learning (ML)-based Network Intrusion Detection System (NIDS) domain and Federated Learning (FL) as they are the foundational concepts for our approach.

\subsection{Machine Learning (ML) based Network Intrusion Detection Systems (NIDS)}\label{sec:mlbasednids}
Network intrusion detection systems (NIDS) are responsible for inspecting all network traffic in a specific device or organization to detect malicious behavior. The NIDS can be categorized regarding aspects of its
granularity and detection method. Regarding granularity, the NIDS inspection can be on each packet during a network communication session. This approach is usually associated with deep packet inspection (DPI). However, this approach has decreased in popularity due to the increase of cryptography in network communications that, in most cases, make it unfeasible to understand the payload data of packets \cite{dpi}. Another level of granularity considers the aggregation of all packets in specific network communication, commonly referred to as network flows. In network flows, the samples usually are aggregations defined by source and destination with four tuples composed of source IP, destination IP, source port, destination port, and 5-tuples, same as the latter but with the addition of the protocol~\cite{flow-based}. The favorable aspect of flow-based NIDS is the reduced amount of data. On the other hand, it can lose some possible particularities just in a packet-level analysis.

The detection method for NIDS can be based on the signature of previous attacks, also known as misuse detection. It represents a high detection rate for known attacks but cannot achieve sound performance for zero-day attacks (previously unknown). The other approach is anomaly-based. It is responsible for determining the expected behavior of the network, and any example that deviates from this expected behavior is classified as anomalous. The anomaly-based NIDS is capable of detecting zero-day attacks. However produces high false positive rates (e.g., unusual network usage or new application in use). 

Additionally, there are two distinct approaches for ML-based IDS, those that consider the learning task as a binary classification between benign (or expected) and malicious (or anomalous) traffic. On the other hand, the multi-class approach that during ML inference aims to classify if a specific test sample is benign or one of the specific attacks under the scope, such as denial of service (DoS), distributed denial of service (DDoS), scanning/probing, among others.

In this work, we consider a binary anomaly flow-based NIDS. Furthermore, the use of machine learning for NIDS is not new. However, the NIDS research domain has specific challenges and pitfalls that must be addressed when evaluating the usage of ML, and we follow these recommendations as possible. For additional discussion and background about these challenges, we suggest other references \citep{theroleofml, outside, dosanddonts}.

\subsection{Federated Learning (FL)}
Federated Learning (FL) is a collaborative learning method for sharing intelligence between multiple parties. First introduced by \cite{pmlr-v54-mcmahan17a} on 2017, it is increasingly being investigated to the NIDS domain~\cite{review_fl_ids_iot_2022}.

\begin{figure}[htb]
    \centering
    \includegraphics{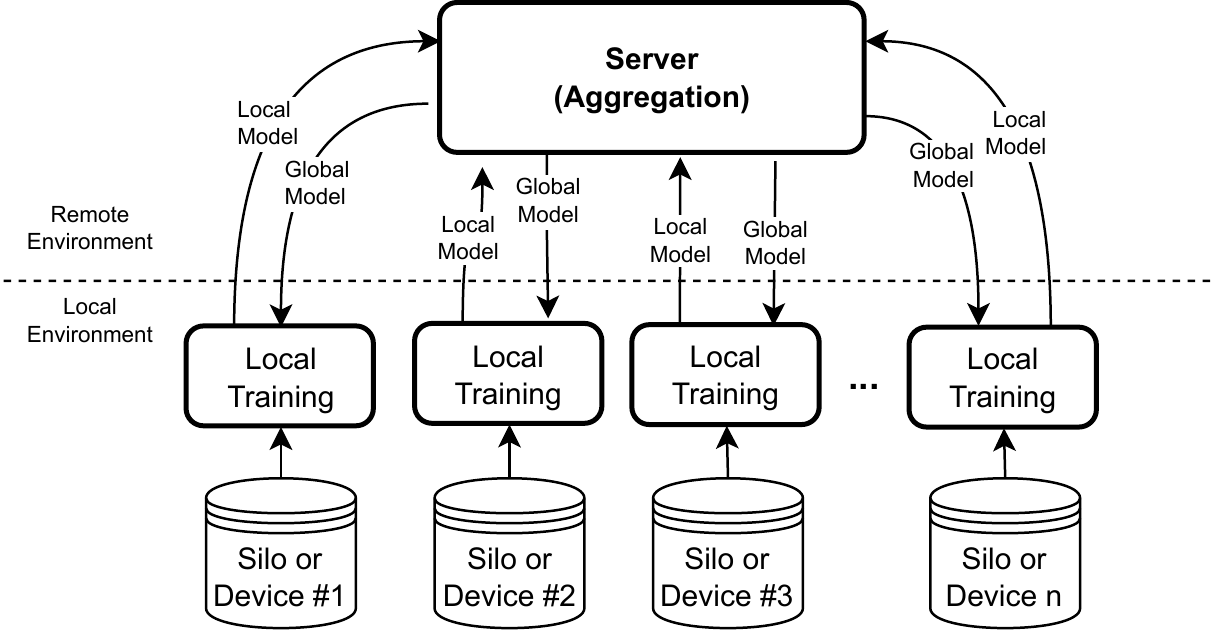}
    \caption{A generic architecture of a Federated Learning setup.}
    \label{fig:federated}
\end{figure}

Figure~\ref{fig:federated} represents a generic architecture of a FL setup. The system comprises many data silos (e.g., an organization) or devices. Initially, these participants retrieve a common global model from a central entity (the server). Then, each participant starts from this global model and performs ML training using its local data. In the next step, these participants share the parameters of their local-trained models (and not their data) with the server. Finally, the server aggregates all received parameters, and the cycle starts again. Each cycle of retrieving a global model, performing local training, and sending the parameters back to the server, is defined as a ``round.''

The internet of things (IoT) context is highly applicable to the FL paradigm given the various edge devices generating data and the need to learn from this vast amount of data~\cite{fl_mobile_survey2020, fl_wireless_2020}. Traditionally, it would require transferring these data to a central server. However, this traditional approach increases network traffic, directly results in latency and consumption of available devices' resources, and raises privacy concerns about sending its data to an external server. Thus, FL enables this learning more securely and efficiently. 

On the FL domain, there are many details, such as the number of parties that could characterize the learning task as cross-silo or cross-device federated learning. In the cross-silo setup, the participants aggregate the data of multiple users, such as an enterprise or an IoT gateway. In the cross-device, devices themself participate in the FL setup. The cross-device setup usually is from hundreds to millions of participants~\cite{fl-book}. The amount of participants and its computational resources introduces many challenges to the FL, such as communication failures and staggering participants. These FL-related challenges are out of our scope.
Also, the horizontal and vertical approaches are essential differences in the FL setup. The horizontal FL is associated with data that shares the same feature set between the participants of the FL. As for NIDS, each participant's data can be made of the same flow-based features (e.g., total bytes) but based on different network sessions. On the other hand, the vertical FL does not share the same feature space between the participants. As in medical or financial data, each participant can have specific attributes of a single person. For instance, an institution can have the credit score of a specific customer and the other institution the purchase history.

The two classic approaches for FL optimization (also referred as aggregation strategy) are the \texttt{FedSGD}~\citep{fedsgd} and \texttt{FedAvg}~\citep{pmlr-v54-mcmahan17a}. With differences in the number of participants, epochs, and batch size for each round, the \texttt{FedAvg} is still the standard baseline for FL. It outperforms the \texttt{FedSGD} due to communication efficiencies. However, both optimizer degrades their performance in data heterogeneity between participants. To overcome this challenge of applying FL to \textit{non-IID} data, the use of regularization techniques have been explored~\citep{ioft}, with prolific research on the area of aggregation strategies for \textit{non-IID} data as the \texttt{FedProx}~\citep{fedprox}, \texttt{FedAvgM}~\citep{FedAvgM}, \texttt{FedOpt} (\texttt{FedAdam}, \texttt{FedAdagrad}, \texttt{FedYogi})~\citep{FedOpt}, and \texttt{Fed+}~\citep{fedplus}.

Our work considers a horizontal cross-silo federated learning setup with four participants (silos). We also explore the effects of \textit{non-IID} data between silos and aggregation strategies.

\section{Related Work}\label{related}
Our work bonds two prolific research domains: the FL-based NIDS and the generalization of ML-based NIDS. Thus, our review analyzes these two domains in this section.
\subsection{Generalization Problem}
The capability to transpose ML-based NIDS beyond its original dataset is an open challenge. \cite{catillo2021, kenyon2020public} reports the challenges of public datasets in representing real network profiles and the inefficiency of generalizing from trained ML models on specific NIDS datasets to real deployments. In this study, we propose federated learning as an effective technique to address this challenge, demonstrating successful performance between diverse datasets compared to the traditional centralized machine learning setup.

In \citep{non_generalize_unsup}, the authors evaluate the capability of unsupervised learning algorithms to generalize between different network contexts. The authors evaluate classic ML algorithms such as isolation forest, local outlier factor, and one-class SVM (oSVM). Also, they evaluate a deep autoencoder for this same task. In contrast to our reported results, they do not report a satisfactory performance in an inter-dataset evaluation (training in one dataset and testing in another). The authors highlight the need for further research in the generalization of unsupervised methods. 
\citep{ongeneralisability} also investigates the generalization capability. The authors evaluate supervised (Extra Tree, Feed Forward, Random Forest, and LSTM) and unsupervised methods (Isolation Forest, oSVM, SGD-oSVM). It reports the best performance of unsupervised methods when compared with supervised. The unsupervised methods have a F1-score decay of $28\%$ when evaluating the trained model on a dataset other than its training dataset. Using two datasets with a common feature set and supervised learning algorithms (classic and deep learning), \cite{ccs-poster} also reports the inability to generalize between two datasets.

Both \cite{crosseval, inter-dataset} proposes methods for evaluating a ML-based NIDS in different datasets. \cite{crosseval} proposes a cross-evaluation method named XeNIDS. The authors propose this method due to the reported fact that \textit{``training an ML-NIDS  on a  (large)  dataset,  such  ML-NIDS  will detect the attacks contained in such dataset''}. So, their method aims to evaluate the proposed ML-based solutions in multiple contexts. \cite{inter-dataset} clearly states that the generalization assumption is false and that further research is needed. Our proposed method aims to provide directions for tackling the generalization challenge.

For generalization in the financial domain, \cite{financialcrime} explains the challenges of data-sharing between financial institutions due to regulations. As previously pointed out, data-sharing is also a challenge for NIDS. The authors then present the application of federated learning for improving the detection of financial crimes between institutions without data-sharing. It reports that the FL-based approach outperforms local trained models by $20\%$.

\subsection{FL-based NIDS}\label{sec:related_flnids}

This section reports previous research using federated learning (FL) for the specific task of network intrusion detection (NIDS). Table~\ref{tab:relatedworks} summarizes all the related work.

In summary, most works that use multiple datasets do not explore the generalization of learning between them~\cite{TIAN2021102344, chenetal2020, qinetal2020, khoa, ferrag2021, zhaoetal2019, dongetal2022}. The common practice adopted by the authors is to evaluate the proposed approach on each dataset in isolation. Just a few discuss the learning generalization as represented by the column ``generalization''~\cite{sarhan2021fl, sunetal2020, similar-DNN-Popoola2021} from Table~\ref{tab:relatedworks}. Our work aligns with the last proposing an FL-based method focusing on generalization between diverse contexts.

One of the challenges in enabling this cross-evaluation between diverse NIDS datasets is the need for standard NIDS features (horizontal FL). Also, most of the FL-based NIDS work with flow-based datasets. The exceptions are those works in the industrial control system (ICS) domain~\cite{HUONG2021103509, lietal2020, mothukuri}. This behavior is due to the datasets currently available in this domain that provides packet-level samples. Other exceptions are the \cite{diot} that basis on a custom dataset developed by the authors using a packet granularity, and the \cite{attota} that uses a dataset with both packet and flow-level features.

Regarding the federated learning setup, most of the related works evaluate just the FedAvg aggregation method. In the case of \cite{CAMPOS2022108661, similar-DNN-Popoola2021}, they relied on the aggregation methods available in the IBM-FL framework~\cite{ibmfl2020ibm}. \cite{chenetal2020} proposed FedAGRU, which, based on FedAvg, uses an attention mechanism for the participants to share or not their parameters to the server.

\newpage
\global\pdfpageattr\expandafter{\the\pdfpageattr/Rotate 90}
\begin{landscape}
\begin{table}[t]
\begin{minipage}[t]{1.2\textwidth}
\centering
\caption{Summary of related works -- Federated Learning-based Network Intrusion Detection Systems.}
\label{tab:relatedworks}
\resizebox{1.2\textwidth}{!}{%
\begin{tabular}{@{}rccccccc@{}}
\toprule
\textbf{Reference}     & \textbf{Dataset}                                                                             & \textbf{Granularity} & \textbf{ML Algorithm}\footnote{NN = Neural Network; DNN = Deep Neural Network; DAE = Deep Autoencoder; VAE = Variational Autoencoder; CNN = Convolutional Neural Network; MLP = Multi-layer Perceptron; DT = Decision Tree; RF = Random Forest; SVM = Support Vector Machine; BNN = Binarized Neural Network; DBN = Deep Belief Network; GBDT = Gradient Boosting Decision Tree; LSTM = Long Short Term Memory; GRU = Gated Recurrent Units}      & \textbf{Aggregation Function}                                               & \textbf{non-IID}                    & \textbf{\begin{tabular}[c]{@{}c@{}}Unsupervised\\ Learning\end{tabular}} & \textbf{Generalization}             \\ \midrule
Rahman et al (2020)~\cite{rahmanetal2020}    & NSL-KDD                                                                                      & Flow                 & NN & FedAvg                                                                      &                                     &                                                                          &                                     \\
Campos et al (2021)~\cite{CAMPOS2022108661}    & ToN\_IoT                                                                                     & Flow                 & Logistic Regression        & FedAvg, Fed+                                                                & \ding{51}          &                                                                          &                                     \\
Sarhan et al (2021)~\cite{sarhan2021fl}    & UNSW-NB15, Bot-IoT                                                                           & Flow                 & DNN and LSTM               & FedAvg                                                                      &                                     &                                                                          & \ding{51}          \\
Tian et al (2021)~\cite{TIAN2021102344}      & CICIDS-2017, IoT-23                                                                          & Flow                 & DAE          & FedAvg                                                                      & \ding{51}          & \ding{51}                                               &                                     \\
Hei et al (2020)~\cite{HEI2020102033}       & KDDCup99                                                                                     & Flow                 & MLP, DT, RF, SVM           & FedAvg                                                                      &                                     &                                                                          &                                     \\
Huong et al (2021)~\cite{HUONG2021103509}     & SCADA                                                                                        & Packet               & VAE-LSTM                   & FedAvg                                                                      &                                     & \ding{51}                                               &                                     \\
Li et al (2020)~\cite{lietal2020}        & SCADA                                                                                        & Packet               & CNN-GRU                    & FedAvg                                                                      &                                     &                                                                          &                                     \\
Mothukuri et al (2021)~\cite{mothukuri} & MODBUS                                                                                       & Packet               & GRU                        & FedAvg                                                                      &                                     &                                                                          &                                     \\
Chen et al (2020)~\cite{chenetal2020}      & \begin{tabular}[c]{@{}c@{}}KDDCup99, CICIDS-2017, \\ WSN-DS\end{tabular}                     & Flow                 & GRU                        & FedAGRU                                                                     & \ding{51}          &                                                                          &                                     \\
Qin et al (2020)~\cite{qinetal2020}       & \begin{tabular}[c]{@{}c@{}}CICIDS2017,\\ ISCX Botnet 2014\end{tabular}                      & Flow                 & BNN               & Majority Vote                                                               &                                     &                                                                          &                                     \\
Khoa et al (2020)~\cite{khoa}      & \begin{tabular}[c]{@{}c@{}}KDDCup99, NSL-KDD, \\ UNSW-NB15, N-BaIoT\end{tabular}             & Flow                 & DBN                        & FedAvg                                                                      &                                     &                                                                          &                                     \\
Attota et al (2021)~\cite{attota}    & MQTT                                                                                         & Packet, Flow               & Ensemble: NN and RF    & FedAvg                                                                      &                                     &                                                                          &                                     \\
Popoola et al (2021)~\cite{popoolazeroday}   & Bot-IoT, N-BaIoT                                                                             & Flow                 & DNN                        & FedAvg                                                                      &                                     &                                                                          &                                     \\
Sun et al (2020)~\cite{sunetal2020}       & \begin{tabular}[c]{@{}c@{}}Custom dataset\\ (LAN-Security Monitoring)\end{tabular}           &      Flow\footnote{The authors use a feature map. Considering as aggregation, we categorize as flow granularity.}                & CNN                        & FedAvg                                                                      &                                     &                                                                          & \ding{51}          \\
Nguyen et al (2019)~\cite{diot}    & Custom dataset (DÏoT)                                                                        & Packet               & GRU                        & FedAvg                                                                      &                                     &                                                                          &                                     \\
Tang et al (2021)~\cite{tang2022}      & CICIDS2017                                                                                   & Flow                 & GRU                        & FedAvg                                                                      &                                     &                                                                          &                                     \\
Ferrag et al (2021)~\cite{ferrag2021}    & MQTTset, BoT-IoT, TON\_IoT                                                                   & Flow                 & DNN, CNN, LSTM             & FedAvg                                                                      & \ding{51}          &                                                                          &                                     \\
Liu et al (2021)~\cite{liuetal2021}       & KDDCup99                                                                                     & Flow                 & MLP                       & FedAvg                                                                      &                                     &                                                                          &                                     \\
Zhao et al (2019)~\cite{zhaoetal2019}      & \begin{tabular}[c]{@{}c@{}}CICIDS2017, ISCXVPN2016, \\ ICSXTor2016\end{tabular}              & Flow                 & DNN                        & FedAvg                                                                      &                                     &                                                                          &                                     \\
Dong et al (2022)~\cite{dongetal2022}      & \begin{tabular}[c]{@{}c@{}}DDoS2019, DoHBrw2020, \\ Darknet2020, Maldroid2020\end{tabular}   & Flow                 & GBDT                       & FedAvg                                                                      &                                     &                                                                          &                                     \\
Popoola et al (2021)~\cite{similar-DNN-Popoola2021}   & \begin{tabular}[c]{@{}c@{}}TON\_IoT, Bot-IoT, \\ CICIDS2018, UNSW-NB15\end{tabular}          & Flow                 & DNN                        & \begin{tabular}[c]{@{}c@{}}FedAvg, Fed+, \\ CM, CM+\end{tabular}            &                                     &                                                                          & \ding{51} \\
\textbf{Our work}      & \textbf{\begin{tabular}[c]{@{}c@{}}TON\_IoT, Bot-IoT, \\ CICIDS2018, UNSW-NB15\end{tabular}} & \textbf{Flow}        & \textbf{\begin{tabular}[c]{@{}c@{}}Stacking \\ Deep Autoencoder\end{tabular}}       & \textbf{\begin{tabular}[c]{@{}c@{}}FedAvg, FedOpt\footnote{FedOpt consists of the following algorithms: FedAdam, FedAdagrad,
FedYogi.}, \\ FedAvgM\end{tabular}} & \ding{51} & \ding{51}                                      & \ding{51} \\ \bottomrule
\end{tabular}%
}
\end{minipage}
\end{table}
\end{landscape}
\global\pdfpageattr\expandafter{\the\pdfpageattr/Rotate 0}

Some related works use sequence models such as Long Short Term Memory (LSTM) and Gated Recurrent Units (GRU) as the ML algorithm. However, the dataset considered in this work is in a network flow granularity. Flow-based datasets do not present time dependency between samples, so sequence models such as LSTM and GRU are not applicable. 
Regarding the learning approach, most previous works use supervised learning. The exceptions are \cite{TIAN2021102344, HUONG2021103509}. Also, the aspects of \textit{non-IID} on federated learning are not evaluated by all previous works, being discussed by \cite{CAMPOS2022108661, TIAN2021102344, chenetal2020, ferrag2021}.

Our work contributes to the body of knowledge by exploring an FL-based NIDS for generalization between heterogeneous networks. In our case, we use an unsupervised method in a stacking setup and evaluate its generalization between heterogeneous networks. We use different datasets to represent these networks. It also evaluates the aspects of \textit{non-IID} (both class and quantity skewness), and we evaluate the proposed method with aggregation algorithms beyond the traditional FedAvg.

\subsubsection{Generalization on NIDS}
In \cite{sarhan2021fl}, federated learning is proposed to share learning (generalization) between two organizations.
The organizations are represented by two distinct datasets (UNSW-NB15 and BoT-IoT) sharing the exact feature set (horizontal FL).
Despite reporting satisfactory results, when evaluating the model trained in one dataset against another, a reduced detection rate of $4.17\%$ and $5.78\%$ was obtained using deep learning and LSTM, respectively. However, there is a lack of discussion related to the difficulty of generalization; their proposal is based on supervised learning.

\cite{similar-DNN-Popoola2021} reports the capability of NIDS generalization through the application of FL, and more specifically, with two aggregation algorithms: the \texttt{Fed+} and the Coordinate Median Plus that is one variation of \texttt{Fed+}. In this work, the authors also use a supervised learning approach, in contrast with our approach using unsupervised learning. Also, their results are tied to two specific aggregation algorithms.

In the direction of being capable of generalizing on the NIDS domain, \cite{EFC} presents the Energy Flow Classifier (EFC) for flow-based NIDS. This algorithm reports the capability of domain adaptation (generalization). Based on the flow features, it calculates the energy associated with each benign flow and given a specific quantile threshold, a specific flow is categorized as benign or malicious, given a calculated cutoff. Regarding domain adaptation, EFC reports better performance than supervised learning algorithms (Naive Bayes, k-Nearest Neighbors, Decision Tree, Support Vector Machine, Multi-layer perceptron, AdaBoost, and Random Forest). Their evaluation consists of training in one dataset and testing in another (the authors used CIC-IDS-2017 and CIC-DDoS-2019). Our work uses the EFC to generate a new flow feature in a stacking ensemble setup for unsupervised learning. Our approach outperforms the EFC detection performance and works unsupervised in contrast with EFC.

\section{Methodology}\label{methodology}
This section details the methodological approach adopted in this work to evaluate the generalization of an intrusion detection system over heterogeneous networks. In our case, heterogeneous networks are the datasets that represent diverse networks. Then, we use an unsupervised learning algorithm and an experimental approach to validate that our proposed architecture improves the generalization between those heterogeneous networks. 

Our design decisions harness the evidence of past results about the generality of this approach as promising~\citep{ongeneralisability}. In addition to the advantages of unsupervised over supervised methods, as the capability to detect previously unknown attacks, as presented in section~\ref{sec:mlbasednids}. Regarding federated learning (FL) NIDS, there is a lack of exploration of unsupervised methods. Most of the previous works do not evaluate the performance of proposed methods in a cross-evaluation scheme, as discussed in section~\ref{sec:related_flnids}. 

\begin{figure}[htb]
    \centering
    \includegraphics{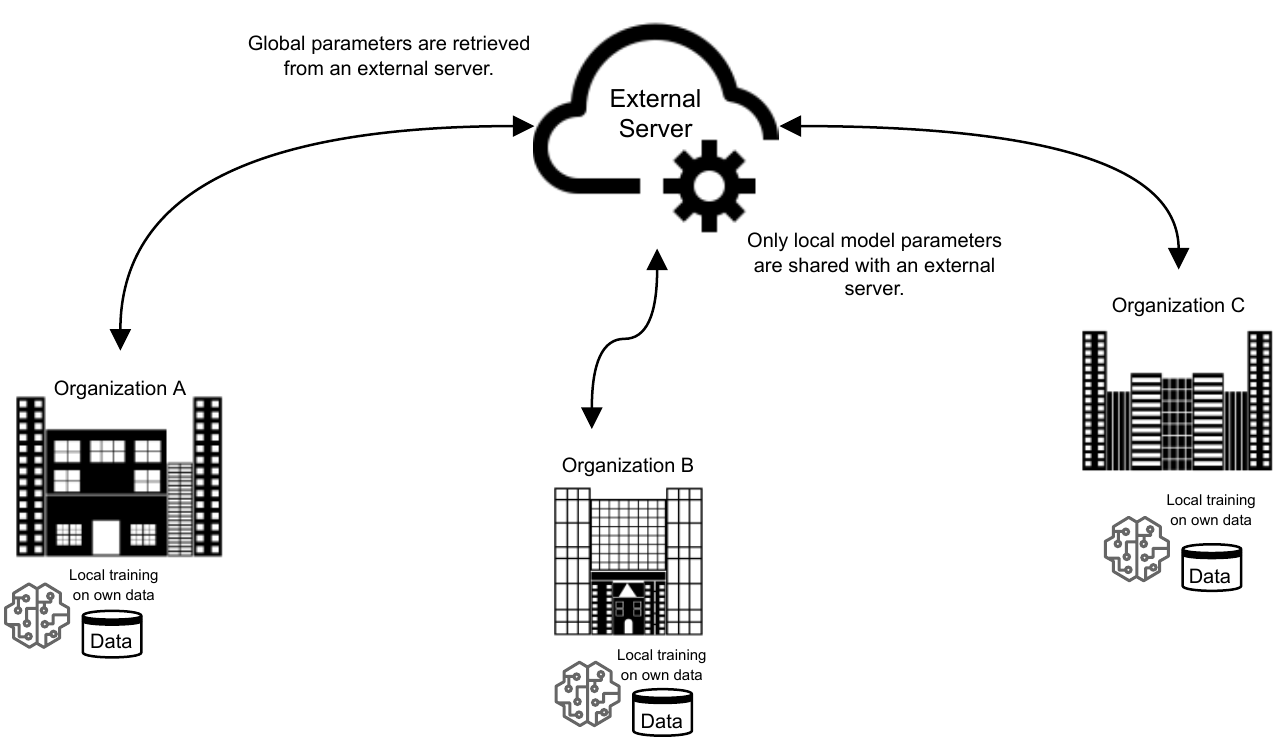}
    \caption{An illustrative environment considering three participants in a federated learning setup. It is composed of multiple organizations sharing only local model parameters. This sharing results in an overall Network Intrusion Detection System (NIDS) solution with contributions learned from data on different environments. Data privacy is achieved.}
    \label{fig:operational}
\end{figure}

The envisioned operational environment under consideration for our work is diverse organizations sharing FL-based NIDS model parameters to compose a more efficient NIDS solution. In this case, the organization can be units of a single entity or various institutions. In both cases, an external server performs the aggregation task. Figure~\ref{fig:operational} presents this operating environment.

\subsection{Data Description}
An essential aspect of our methodology is that all participants of the FL share the same feature set. This design decision is tied with the requirement for a horizontal FL setup. For this reason, we choose four available datasets with a feature set obtained with the NetFlow~\cite{standardfeatureset2021}.

These four datasets consist of 43 NetFlow features (flow-based). Based on a five-tuple composed of IP source and destination addresses, source and destination ports, and protocol. Each flow (sample) comprises defining features, such as flow duration, number of bytes, and packets. In these datasets, they are all numerical features. Despite this custom version based on the NetFlow features, follow a brief description of each dataset.

\subsubsection*{UNSW-NB15}
It is a network-based dataset that was created to substitute the old datasets used in the NIDS research (KDD-Cup99 and NSL-KDD). It introduces more recent network behaviors and attacks. The dataset was generated in a testbed using a traffic generator capable to generated both attack and normal traffic. Its original 49 flow-based features were obtained with the Argus\footnote{Argus: \url{https://openargus.org/}} and Bro-IDS\footnote{Bro-IDS is now Zeek: \url{https://zeek.org/}}. The attacks present on this dataset are analysis, backdoors, denial of service (DoS), exploits, generic, reconnaissance, shellcode, and worms. Details about this dataset are available in its publication~\cite{unsw-nb15}.

\subsubsection*{CSE-CIC-IDS-2018}
The authors also claiming the lack of adequate datasets. By the date of dataset release, the authors evaluated 11 datasets and claimed they were outdated and unreliable to use. Thus, the authors published a dataset composed of normal and attacks as network flows. The attacks that compose the dataset are DoS, Heartbleed, SSH-Patator, FTP-Patator, Web attack, infiltration, bot, portscan, and distributed DoS (DDoS). The dataset was generated in a testbed divided into two networks (victims and attackers). The original dataset is available with 80 flow features extracted with the CICFlowMeter\footnote{CICFlowMeter: \url{https://github.com/ahlashkari/CICFlowMeter}}. More details about this dataset are available in its publication~\cite{csecicids2018}.
\subsubsection*{Bot-IoT}
The Bot-IoT is a dataset that considers the internet of things (IoT) context. Thus, this dataset addresses the lack of a dataset with botnet samples. The testbed is based on simulated IoT services using MQTT protocol and based on virtual machines (VM). In this dataset, the authors use the Argus to extract network features and also propose other using a sliding window. The Ostinato tool generates normal traffic, and Kali VMs are responsible for generating the attacks (service scanning, OS fingerprint, DoS, DDoS, keylogging, and data theft). A characteristic of this dataset is a high imbalance in favor of attack samples. More details about this dataset are available in its publication~\cite{botiot}.
\subsubsection*{ToN-IoT} 
It is also a IoT-focused dataset. In addition, the authors considers Industrial IoT (IIoT), and their testbed provides a more complete IoT architecture. This testbed considers three layers of devices, the cloud layer, the fog layer (using virtualization), and an edge layer composed with physical devices. The attacks present in this dataset are scanning, DoS, ransomware, backdoor, injection attack, cross-site scripting (XSS), password cracking, and man-in-the-middle (MITM). The dataset originally provides custom features for the IoT devices, host-based features for Windows and Linux datasets. And regarding network, it provides a version with 44 features extracted with Zeek. More details about this dataset are available in its publication~\cite{toniot_network_features}.

A highlight of these datasets is that they are composed of various attacks and regular traffic. A solution capable of generalizing between them can be advantageous to the participants once they can anticipate detecting attacks never previously seen in their network context.

\subsection{Proposed Architecture}\label{sec:architecture}
\begin{figure*}[htb]
    \centering
    \includegraphics[width=\textwidth]{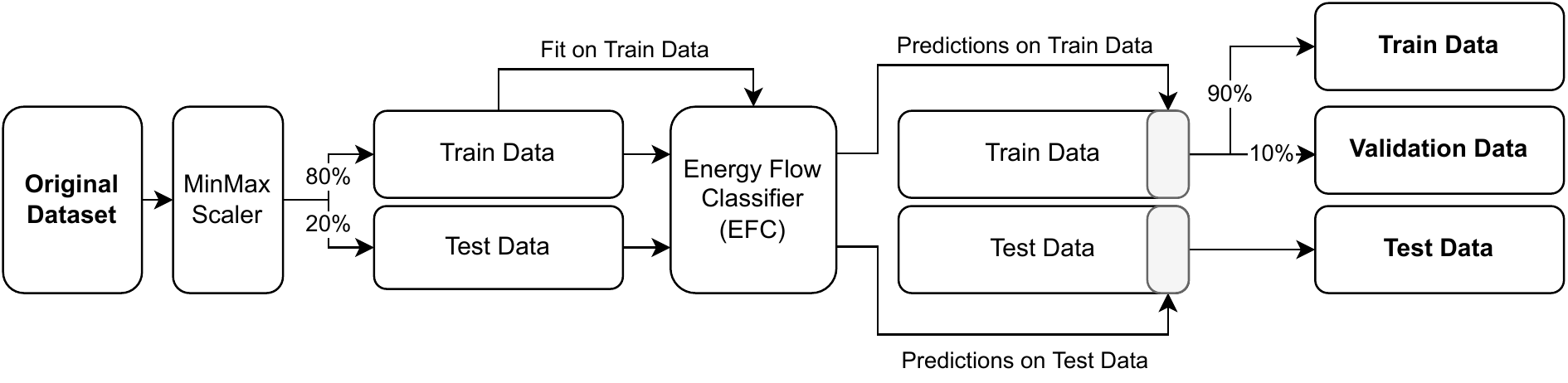}
    \caption{The overall setup for the proposed stacked-unsupervised learning setup. It reports the overall data processing, from the original datasets, the scaling step with MinMax, the split of data, the usage of EFC as part of the stacking as new feature for the original dataset, and the final split of the data.}
    \label{fig:datasetprocessing}
\end{figure*}

Figure~\ref{fig:datasetprocessing} presents our proposed architecture. In our methodology, we initially perform a scaling step from each of the original datasets (min-max scaler) before using each dataset. Then, the dataset split between train and test sets are 80\% and 20\%, respectively.

Regarding stacking, we use the Energy Flow Classifier (EFC) as a first step in the overall architecture. The decision to use EFC as part of our architecture is that it is the only algorithm with the proven capability of generalization between datasets to the best of our knowledge. Then, we train an EFC only on training data, and the predictions obtained on both train and test sets are appended to the original sets. Later, the train data with EFC predictions are split on 90\% from training and 10\% for validation (during autoencoder training). The test data remains 20\% of the original data.

\subsection{Deep Autoencoder}\label{sec:deepautoencoder}
\begin{figure}
    \centering
    \includegraphics[width=\columnwidth]{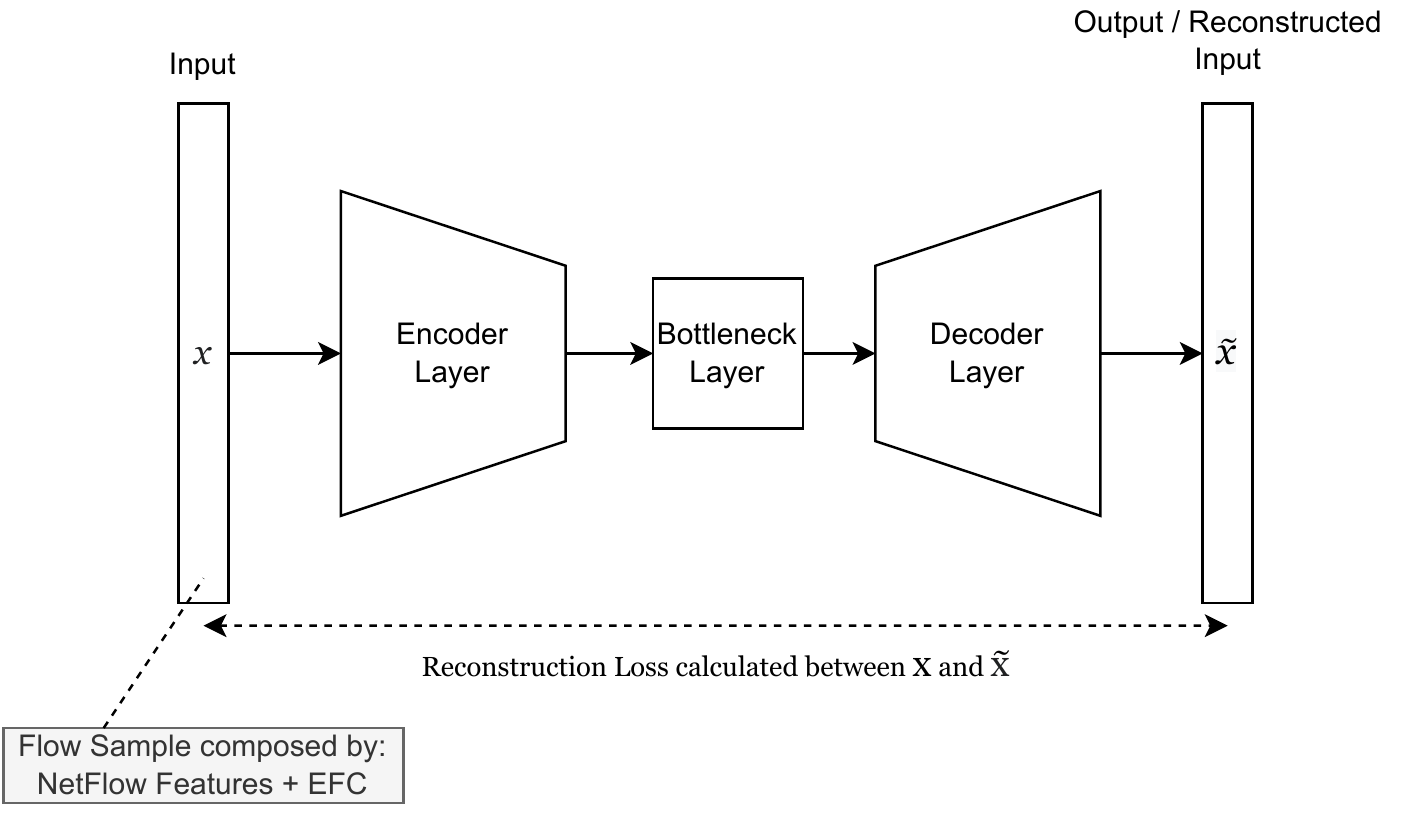}
    \caption{Deep Autoencoder architecture composed by input ($x$), encoder layer, bottleneck layer, decoder layer and output or reconstructed input ($\tilde{x}$). The reconstruction loss is calculated between $x$ and $\tilde{x}$.}
    \label{fig:dae_architecture}
\end{figure}

The unsupervised learning algorithm adopted by us is a deep autoencoder that has been applicable to problems such as image denoising, dimensionality reduction, and anomaly detection. The autoencoder ($f$) is responsible for given an input ($x$) generate a output ($f(x)=\tilde{x}$) that is similar to its input ($\tilde{x} \approx x$). The structure of autoencoders is composed of three layers, an encoder network, a bottleneck section also known as latent space or compressed representation, and a decoder network. All three layers are neural networks. This overall architecture of the autoencoder is detailed by Figure~\ref{fig:dae_architecture}.

The process of training a deep autoencoder aims to minimizes the reconstruction loss ($\mathcal{L}(x,\tilde{x})$), from a given input $x \in \mathbb{R}^N$ minimizes the distance between the input to the output $\tilde{x} \in \mathbb{R}^N$.  In our case, $N$ represents the number of flow-based features associated with each sample $x$ from the datasets. 

During the training of our autoencoder, the input are the training data composed by the flow-based features and the EFC predictions. The output used for distance calculation during training is the same as the input data. During this training process we uses the Mean Squared Error (MSE) as the loss function. The parameters of the autoencoder that we use in this work is summarized in Table~\ref{tab:deep-autoencoder}. 

\begin{table}[htb]
\centering
\caption{Deep Autoencoder Architecture and Hyperparameters}
\label{tab:deep-autoencoder}
\resizebox{0.6\textwidth}{!}{%
\begin{tabular}{@{}ll@{}}
\toprule
\textbf{Attribute}                & \textbf{Value}     \\ \midrule
Input / Output Layer (\# neurons) & Variable                 \\
Encoder Layer (\# neurons)        & 32 : 16 : 8        \\
Bottleneck Layer (\# neurons)     & 4                  \\
Decoder Layer (\# neurons)        & 8 : 16 : 32        \\
Hidden Layers Activation Function & ReLU               \\
Output Layer Activation Function  & Sigmoid            \\
Optmization Algorithm             & Adam               \\
Training Loss Function                     & Mean Squared Error \\
Learning Rate                     & 0.001              \\
Batch Size                        & 128                \\
Epochs                            & 10                 \\ \bottomrule
\end{tabular}%
}
\end{table}

The input and output layers from Table~\ref{tab:deep-autoencoder} is reported as a variable because using EFC as a feature can be removed. Also, from the original datasets with 42 features, the following features are removed: IP source and destination addresses and source and destination ports. This decision avoids learning the specific characteristics of the testbed for dataset generation (e.g., specific IP generated specific attack traffic) as also pointed out by~\citep{troubleshoot_cicids} regarding shortcut learning. Removing IPs and ports aims to make the model learn the flow properties instead of the particular endpoints responsible for each traffic. Because of being an unsupervised learning task, the feature attack that represents each specific attack contained on the datasets, the feature dataset representing the dataset name, and the feature label with 1 for attack and 0 for normal flow are removed. The features list and description are available in~\citep{standardfeatureset2021}.

\subsubsection{Detecting Attacks with Autoencoder}
After the training of the Autoencoder, we calculate a threshold for evaluating the test data. If a test sample presents a reconstruction loss greater than this threshold, then this sample is considered an attack. In our case, the reconstruction loss under consideration during threshold calculation and inference is the mean absolute error (MAE):

$$
\text{Reconstruction Loss} = \text{MAE} = \frac{1}{N}\sum_{i=1}^{N} |x_i -\tilde{x}_i|
$$

Our work proposes using two possible thresholds for each client participating in the federated learning scheme. 
These thresholds are calculated locally on each client (i.e., silo instance). Before the federated learning execution, we separate $10\%$ from the training data to create a validation set (also represented in Figure~\ref{fig:datasetprocessing}). Then this validation data are separated locally on only benign 
or only attack samples, respectively, $\mathcal{B}=\{(x_m, \text{Benign})\}$ and $\mathcal{A}=\{(x_n, \text{Attack})\}$. $m$ represents the number of benign samples, $n$ is the number of attack samples, and $m+n$ is the total samples of the validation set.

In each round, during local evaluations, we use these two validation subsets ($\mathcal{B}$ and $\mathcal{A}$) to recalculate the reconstruction loss for both benign and attack samples. The definition if a test sample is benign or attack uses the benign ($T_\mathcal{B}$) and attack ($T_\mathcal{A}$) thresholds, they are the MAE calculated over the validation subsets and the local inference performed by the autoencoder ($\tilde{\mathcal{B}}=\{\tilde{x}_{m}\}$ and $\tilde{\mathcal{A}}=\{\tilde{x}_{n}\}$): 
$$
T_\mathcal{B} = \text{MAE}(\mathcal{B}, \tilde{\mathcal{B}}), \hspace{5mm}
T_\mathcal{A} = \text{MAE}(\mathcal{A}, \tilde{\mathcal{A}}) .
$$

For the inference of a sample using both thresholds, the sample calculated loss ($\mathcal{S}_{loss}$) is considered an attack if $|\mathcal{S}_{loss} - T_\mathcal{B}| > |\mathcal{S}_{loss} - T_\mathcal{A}|$, otherwise it is considered benign. For the case of only benign threshold, if $\mathcal{S}_{loss} > T_\mathcal{B}$, then it is considered an attack. Otherwise, it is considered benign. A point of attention for this two thresholds approach is to deploy the solution. All the silos must have an initial training dataset composed of benign and attack samples. Using only a benign threshold is still possible and evaluated in our work.

\subsection{Federated Learning}
From the augmented data obtained with the stacking strategy detailed in section~\ref{sec:architecture}, and the deep autoencoder presented in section~\ref{sec:deepautoencoder}, in this section, we present the federated learning setup.

Each dataset represents a silo in our cross-silo federated learning setup. Thus, we have four clients. Each client is composed of the dataset and the autoencoder architecture. In the initialization of the setup, each client retrieves the model's architecture from the global server with random weights.

Then, for each round, the client performs the training on its local data using ten epochs and a batch size of 128. After this training, the clients may perform a local evaluation (test the trained model on its test set) or directly send its model weights to the server for aggregation. The aggregation may be performed with many algorithms such as FedAvg and FedOpt. In our case, the use of different aggregation algorithms is reported in section~\ref{results}.

In this FL setup, many aspects can be evaluated, such as the impact of staggering participants, impacts of communication latency, trustworthy behavior of clients, and adversarial attacks, among others. However, these FL-related challenges are not evaluated in this work.

\subsection{Evaluation Criteria}\label{sec:evaluation_criteria}
To evaluate our proposed method and the baselines, we consider the F1-score. The F1-score is a valuable learning metric for unbalanced data. Unbalanced data is inherently the case of network intrusion detection, wherein in the typical environment, the majority of samples in composed of regular traffic and a minority of malicious samples.

The F1-score is a harmonic mean of the precision and recall (also known as true positive rate - TPR). The precision metric measures the proportion of positive inferences that are actually correct. In comparison, recall measures the proportion of actual positives that are identified correctly. These metrics can be easily interpretable from the traditional confusion matrix, as presented in Table~\ref{tab:confusion}. In our case, the positive class is the ``attack.'' For instance, a flow-based sample from ``normal'' traffic, if inferred by our method as an ``attack,'' would count as a false positive (FP). However, if predicted as ``normal,'' it counts as a true negative (TN).

\begin{table}[htb]
\centering
\caption{A confusion matrix for the classification of network flow-based samples. The predict columns refers to the output of the inference based on the input samples.}
\label{tab:confusion}
\resizebox{\textwidth}{!}{%
\begin{tabular}{@{}rcc@{}}
\toprule
                            & \multicolumn{1}{l}{\textbf{Predict as ``Attack''}} & \multicolumn{1}{l}{\textbf{Predict as ``Normal''}} \\ \midrule
\textbf{Sample is ``Attack''} & True Positive (TP)                               & False Negative (FN)                              \\
\textbf{Sample is ``Normal''} & False Positive (FP)                              & True Negative (TN)                               \\ \bottomrule
\end{tabular}%
}
\end{table}

Formalizing the concepts of F1-score, Precision, and Recall in terms of the parameters from the confusion matrix, results in the equations:

$$
    \text{Precision}=\frac{TP}{TP+FP}\,, \hspace{5mm}
    \text{Recall}=\frac{TP}{TP+FN}\,,
$$
$$
    \text{F1-score}=2\times\frac{\text{Recall}\times\text{Precision}}{\text{Recall}+\text{Precision}}\,\text{.}
$$

When evaluating the performance of each silo after the end of the federated learning simulation, we also report additional learning metrics to support the interpretability of these results. The additional metric is the accuracy, which, despite being misleading on unbalanced cases, is being reported for completeness. The other metrics are more related to the needs of the intrusion detection problem. The Missrate, also known as false negative rate,  measures the attacks that have been missed by the system (false negatives - FN) and are an essential metric due to the fact of the high cost of errors in intrusion detection. A simple missed sample can represent the compromise of the system. This high cost of error is not present in other ML domains such as natural language processing or product recommendation~\cite{outside}. The fallout metric, also known as false positive rate, is concerned about false positives (FP) once it induces high human intervention to filter out false alarms. The last metric is the area under the ROC curve (AUC). The receiver operating characteristic curve (ROC) measures the performance of a model in terms of true positive (TPR) and false positive rates (FPR) for multiple decision thresholds. The AUC aggregates ROC. It measures the model's quality irrespective of what classification threshold is chosen.

Based on Table~\ref{tab:confusion}, these metrics can be represented by the following equations:

$$
\text{Accuracy} = \frac{TP+TN}{TP+TN+FP+FN},\;\;
\text{Missrate} = \frac{FN}{FN+TN},
$$
$$
\text{Fallout} = \frac{FP}{FP+TN}.
$$

\subsection{Baselines}\label{sec:method_baselines}
For comparison of our proposed method with other approaches, we choose as baselines the Energy Flow Classifier (EFC) algorithm that was previously proven to generalize between datasets~\citep{EFC}. Also, the classical unsupervised algorithms Isolation Forest (IF) and Local Outlier Factor (LOF), as recommended by \cite{dosanddonts}, address the pitfall for inappropriate baselines using simpler methods. Additionally, we evaluate a deep autoencoder as the chosen approach by \cite{non_generalize_unsup, TIAN2021102344}. 

Regarding the FL-based NIDS listed in Table~\ref{tab:relatedworks}, only the unsupervised methods are under consideration. In our case, the deep autoencoder is similar to \citep{TIAN2021102344}. The sequence model proposed by \citep{HUONG2021103509} (LSTM) does not apply to flow-based data. Thus, we also evaluate the deep autoencoder in an FL setup.

For the classical unsupervised algorithms, the default hyperparameters are in use. Just for the deep autoencoder that we considered the architecture reported in section~\ref{sec:deepautoencoder} and ten training epochs. For the autoencoder in naive evaluation (local training and cross-silo evaluation), the reconstruction loss was obtained from the mean absolute error over the validation dataset and using the loss value from the 95\% quantile of the reconstruction losses. 

\section{Results and Discussion}\label{results}
This section reports the results obtained with the methodology presented in section~\ref{methodology}. In addition, the results are presented with the respective discussion to facilitate the interpretation and derivations from them and to understand the adopted methodological decisions.

The results were achieved with a CPU Intel(R) Xeon(R) Gold 5117 with 14 cores and 128 GB RAM, using the Tensorflow $2.9.1$, Scikit-Learn $1.1.1$, and the federated learning framework Flower $0.19$. The reproducible code is available on our laboratory's repository\footnote{Repository: \url{https://github.com/c2dc/fl-unsup-nids}}.

\subsection{Silos (Datasets) Configurations -- \textit{non-IID} influence}
The three datasets configuration for evaluating our method are the original datasets, a sampled version containing 1 million samples of these original datasets, and a reduced version. In our case, each dataset is a data silo for the federated learning setup. A summary of the number of samples and the binary class (benign or attack) distribution is reported in Table~\ref{tab:descriptive_datasets}. These three dataset configurations are based on past research references. From Table~\ref{tab:descriptive_datasets} we can confirm that despite the TON-IoT reduced version, in all other cases, the class distribution is the same, despite the differences in the number of samples.

\begin{table*}[htb]
\centering
\caption{Datasets distribution (number of samples) for the three versions evaluated. The original datasets, the sampled version, with 1 million samples from original dataset, and the reduced version of the dataset.}
\label{tab:descriptive_datasets}
\resizebox{\textwidth}{!}{%
\begin{tabular}{@{}rcccccc@{}}
\toprule
\multicolumn{1}{l}{}              & \multicolumn{2}{c}{\textbf{Original}}   & \multicolumn{2}{c}{\textbf{Sampled}} & \multicolumn{2}{c}{\textbf{Reduced}} \\ \midrule
\textbf{Dataset}                  & \textbf{Benign}   & \textbf{Attack}     & \textbf{Benign}  & \textbf{Attack}   & \textbf{Benign}  & \textbf{Attack}   \\
\multirow{2}{*}{Bot-IoT}          & 135.037 (0.4\%)   & 37.628.460 (99.6\%) & 3.569 (0.4\%)    & 996.431 (99.6\%)  & 1.285 (0.5\%)    & 277.496 (99.5\%)  \\
                                  & \multicolumn{2}{c}{37.763.497}          & \multicolumn{2}{c}{1.000.000}        & \multicolumn{2}{c}{278.781}          \\
\multirow{2}{*}{TON-IoT}          & 6.099.469 (36\%)  & 10.841.027 (64\%)   & 359.618 (36\%)   & 640.382 (64\%)    & 54.024 (25\%)    & 159.436 (75\%)    \\
                                  & \multicolumn{2}{c}{16.940.496}          & \multicolumn{2}{c}{1.000.000}        & \multicolumn{2}{c}{213.460}          \\
\multirow{2}{*}{UNSW-NB15}        & 2.295.222 (96\%)  & 95.053 (4\%)        & 960.078 (96\%)   & 39.922 (4\%)      & 455.751 (97\%)   & 11.788 (3\%)      \\
                                  & \multicolumn{2}{c}{2.390.275}           & \multicolumn{2}{c}{1.000.000}        & \multicolumn{2}{c}{467.539}          \\
\multirow{2}{*}{CSE-CIC-IDS-2018} & 16.635.567 (88\%) & 2.258.141 (12\%)    & 880.623 (88\%)   & 119.377 (12\%)    & 154.649 (87\%)   & 22.490 (13\%)     \\
                                  & \multicolumn{2}{c}{18.893.708}          & \multicolumn{2}{c}{1.000.000}        & \multicolumn{2}{c}{177.139}          \\ \bottomrule
\end{tabular}%
}
\end{table*}

\begin{figure}[]
    \centering
    \begin{subfigure}{\textwidth}
        \includegraphics[width=\columnwidth]{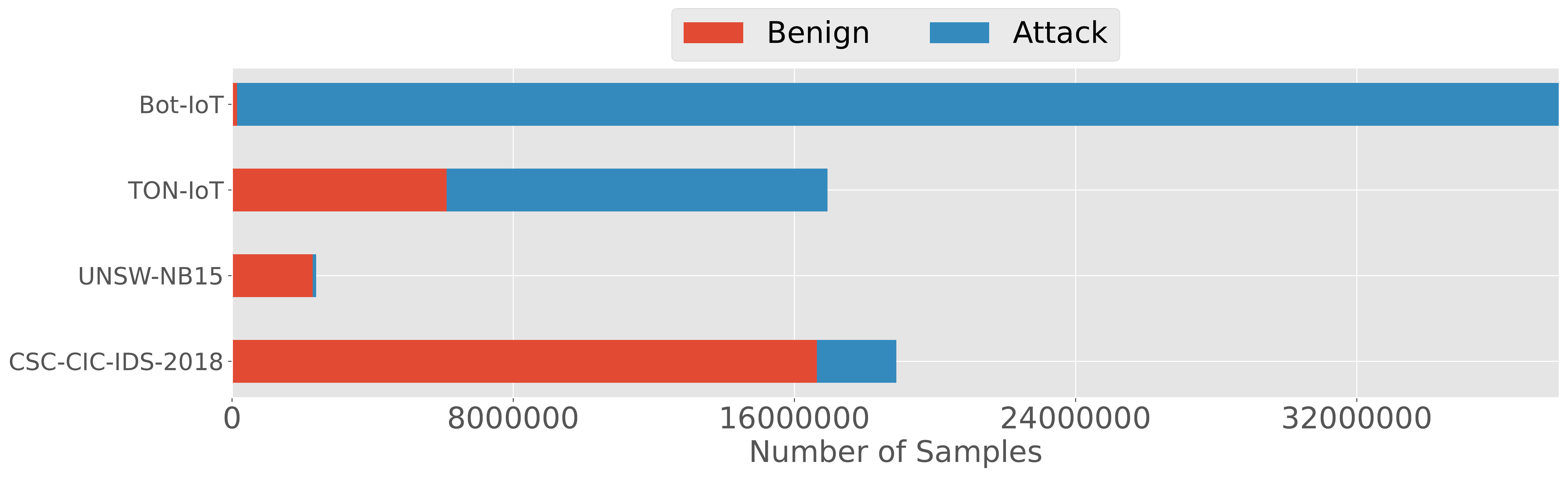}
        \caption{The number of data samples from the original datasets \citep{standardfeatureset2021}, considering benign and attack samples. The quantity and class skewness are present.}
        \label{fig:original_dataset}
    \end{subfigure}
    \hfill
    \begin{subfigure}{\textwidth}
        \includegraphics[width=\columnwidth]{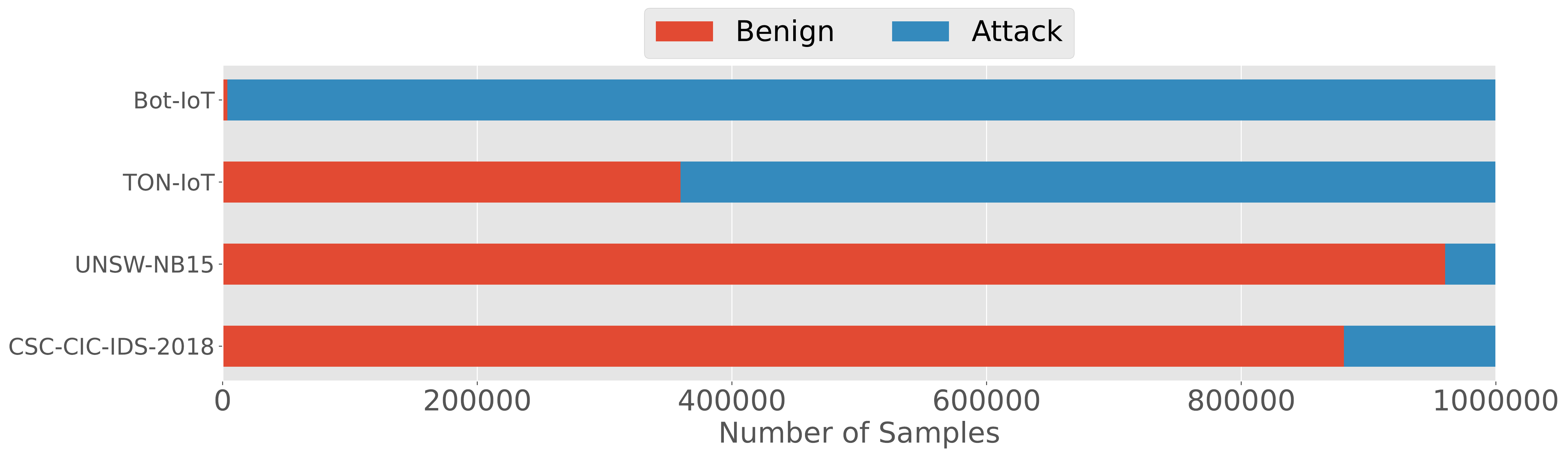}
        \caption{The number of data samples from the sampled datasets (1 million samples) as reported by \cite{ongeneralisability}, considering benign and attack samples. The class skewness is present.}
        \label{fig:1mm_dataset}
    \end{subfigure}
    \hfill
    \begin{subfigure}{\textwidth}
        \centering
        \includegraphics[width=\columnwidth]{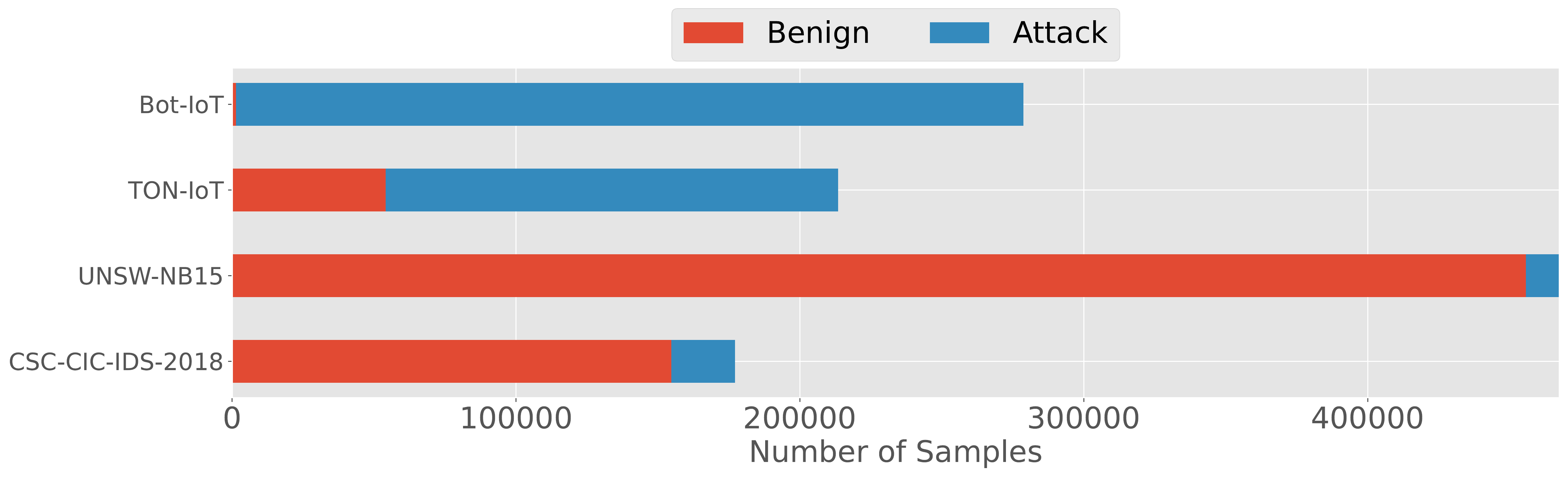}
        \caption{The number of data samples reported by \cite{similar-DNN-Popoola2021} referred as ``reduced,'' considering benign and attack samples. The quantity and class skewness are present.}
        \label{fig:bench_dataset}
    \end{subfigure}

    \caption{Three different data silos configurations to analyze the \textit{non-IID} influence on the federated learning.}
    \label{fig:figures}
\end{figure}

The skewness of the data silos is an essential factor in a federated learning setup. Figure~\ref{fig:figures} supports the visual evaluation of the quantity and class skewness for each configuration. From these figures, the sampled version of the datasets (Figure~\ref{fig:1mm_dataset}) does not present a quantity skewness compared to the other two configurations. Nevertheless, the class skewness is still present.

We first evaluate the three configurations (original\footnote{For the simulation with the original datasets, due to computational resource constraint (i.e., RAM), a feasible approach was to reduce just the Bot-IoT by 50\%, from 37.763.497 to 18.893.708 samples, this is the same quantity as CSE-CIC-IDS-2018 (the second largest dataset).}, sampled, reduced) in a cross-silo configuration for ten rounds with just the FedAvg aggregation technique. There is a minor difference in the average F1-score achieved during tests on the 10\textsuperscript{th} round (original=$0.841$, sampled=$0.842$, reduced=$0.853$). Thus, to further investigate the other variables in our setup, we move along with just the sampled configuration because it requires less computational resources in comparison with the original dataset and provides more samples than the reduced configuration. 
The sampled and reduced configurations presented similar performance from the previous analysis and, as discussed next, for baseline performance. Thus, in our case, the quantity skewness was not a significant factor of influence. Thus, the reported results in the following subsections are mainly based on the sampled silos configuration.

\subsection{Federated Learning Strategies}\label{sec:strategies}
An initial evaluation is the influence of different federated learning strategies (also known as fusion or aggregation algorithms) for the learning task. As available by the Flower framework, we simulated a ten rounds setup with the following strategies: FedAvg, FedAvgM, FedAdagrad, FedYogi, and FedAdam with the default hyperparameters. 

\begin{figure*}[htb]
    \centering
    \includegraphics[width=\textwidth]{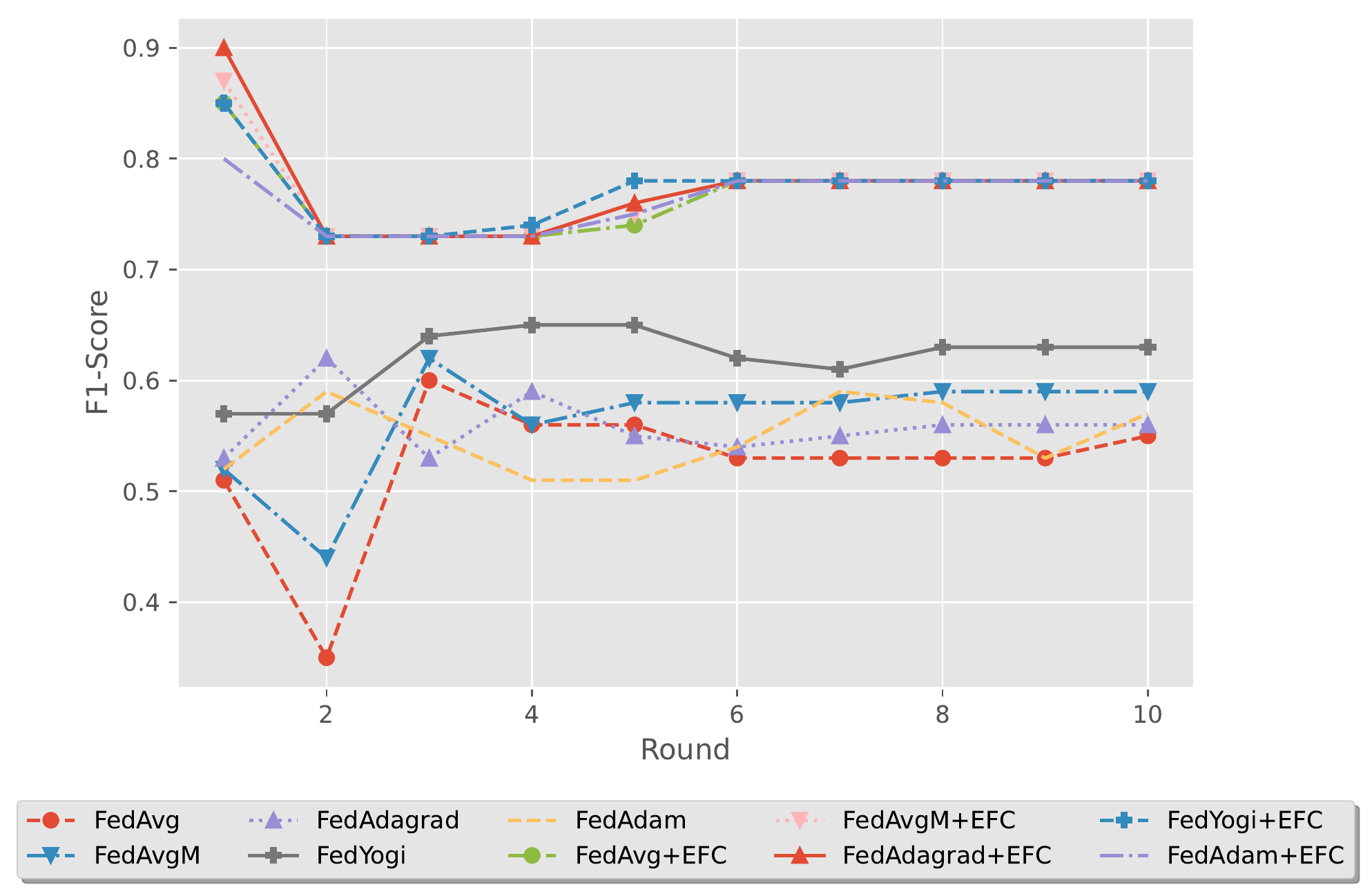}
    \caption{Evaluation of the Federated Learning Strategy in use (FedAvg, FedAvgM, FedAdagrad, FedYogi, FedAdam) and the F1-score performance for 10 rounds. The performance for the Deep Autoencoder setup with and without the EFC algorithm in the stacking setup.}
    \label{fig:strategies}
\end{figure*}

The results are summarized in Figure~\ref{fig:strategies}. When the EFC is not an additional feature as originally proposed, we observed a fluctuation of the F1-score metric during the 10 rounds. The worst case is the FedAdam strategy oscillating over all rounds, and the FedYogi is the best performer, achieving a F1-score of $0.63$ at the 10\textsuperscript{th} round.

The initial fluctuation on the first two rounds with and without EFC reported in Figure~\ref{fig:strategies} is because, on the initial rounds of the federated learning simulation, the server was not waiting for all silos to be online prior to starting. From round 3 and on, all the silos are online and participating in the federated learning setup.

\subsection{Stacking Approach}
A contribution proposed by us is using an autoencoder stacked with the EFC (Figure~\ref{fig:datasetprocessing}). To measure the contribution of EFC to a federated learning setup using the deep autoencoder, we performed an analysis simulating the federated learning without the stacking of EFC. 

For both sampled and reduced datasets using the FedAvg strategy for ten rounds, the average F1-score on the 10\textsuperscript{th} round were $0.46$ and $0.49$, respectively. Analyzing each silos' performance, we confirmed that UNSW-NB15 and CSE-CIC-IDS-2018 were the worst performers without the EFC, pulling the average F1-Score down.

Using the EFC as part of the stacked architecture, the strategy did not demonstrate influence on the federated learning (Section~\ref{sec:strategies}), being a robust setup despite the strategy in use. For all strategies, the F1-score on the 10\textsuperscript{th} round is $0.78$ when considering a reconstruction loss calculation based only on benign samples.

Furthermore, we can infer that EFC is a vital feature to be incorporated into ML-based NIDS using a stacking strategy. It also highlights the importance of further investigating descriptive features for flow-based network samples and their contribution to improving generalization in addition to the federated learning approach.

\subsection{Comparison with Baselines}
The next step was evaluating our proposed method in contrast to the baselines from section~\ref{sec:method_baselines}. The baselines are evaluated locally, with the training and testing sets derived from the same silo. Next, the baselines are evaluated on a cross-evaluation setup, training on the original dataset silo but testing on the test set from the other silos. For instance, training on Bot-IoT, testing on TON-IoT, UNSW-NB15, and CSE-CIC-IDS-2018. The cross-evaluation is then summarized as average F1-scores. The local evaluation is summarized on an average F1-score achieved on each of the four silos for each baseline. The results are summarized on the Table~\ref{tab:f1-results}.

\begin{table*}[htb]
\centering
\caption{Evaluation of baselines and proposed method on the four datasets/silos (sampled version). The average local F1-score represents the average from each dataset. The average cross F1-score the average from all cross-evaluations. Our proposal and the Autoencoder without (w/o) EFC reports the F1-score on the 10\textsuperscript{th} round of the FL setup.}
\label{tab:f1-results}
\resizebox{\textwidth}{!}{%
\begin{tabular}{@{}rcc@{}}
\toprule
\textbf{Algorithm}                       & \textbf{Average Local F1-score} & \textbf{Average Cross F1-score} \\ \midrule
Energy Flow Classifier (EFC)             & $0.77 \pm 0.14$                           & $0.52 \pm 0.38$                            \\
Isolation Forest (IF)                        & $0.61 \pm 0.38$                           & $0.18 \pm 0.24$                            \\
Local Outlier Factor (LOF)               & $0.66  \pm 0.38$                          & $0.03  \pm 0.05$                          \\
Deep Autoencoder (q=0.95, epochs=10)       & $0.35 \pm 0.26$                           & $0.41 \pm 0.34$                           \\ \midrule
\textbf{Federated Learning}& \multicolumn{2}{c}{\textbf{F1-score @ 10\textsuperscript{th} round}}
\\ \midrule
Deep Autoencoder (only benign threshold) & \multicolumn{2}{c}{$0.51$}                                          \\
Deep Autoencoder (benign and attack threshold) & \multicolumn{2}{c}{$0.47$}                                          \\
Our proposal (only benign threshold)       & \multicolumn{2}{c}{$0.77$}                                          \\ 
Our proposal (benign and attack threshold) & \multicolumn{2}{c}{$0.84$}                                          \\

\bottomrule
\end{tabular}%
}
\end{table*}

We can confirm the best local performance for Energy Flow Classifier (EFC) by analyzing the results from Table~\ref{tab:f1-results}. Also, the EFC presented a lower standard deviation representing a more stable performance over the four silos in a localized evaluation setup. By contrast, both Isolation Forest (IF) and Local Outlier Factor (LOF) presented high standard deviations. For instance, IF presented a F1-score of $0.95$ on the sampled version of UNSW-NB15 but $0.01$ on the sampled version of the Bot-IoT ($0.58$ for the sampled TON-IoT, and $0.92$ for CSE-CIC-IDS-2018). Similarly, the LOF presented a high F1-score for the sampled version of UNSW-NB15 ($0.97$) and a low F1-score of $0.01$ for the Bot-IoT ($0.72$ for sampled TON-IoT and $0.93$ for sampled CSE-CIC-IDS-2018). 

This performance behavior based on the localized evaluations allows us to conclude the impact of \textit{non-IID} ($99\%$ of attack samples on Bot-IoT, Table~\ref{tab:descriptive_datasets}) on the LOF and IF algorithms. The local evaluation on the reduced version of the datasets, as expected, is close to the average F1-scores obtained on sampled versions, also with a high standard deviation for LOF ($0.62 \pm 0.37$), IF ($0.60 \pm 0.38$), and deep autoencoder ($0.51 \pm 0.23$). Moreover, the best local performance kept with EFC, achieving an average local F1-score of $0.78 \pm 0.16$.  

The main focus of our proposed method is to achieve good performance in overall silos (heterogeneous networks), generalizing the learning. Therefore, in the cross-evaluation of the baseline algorithms, the best performer is the EFC with an average cross F1-score of $0.52 \pm 0.38$. The classic unsupervised algorithms presented a lower performance on this cross-evaluation.
The cross-evaluation on the reduced version of the datasets, as expected, were similar to the sampled case for average cross F1-scores: EFC ($0.53 \pm 0.40$), IF ($0.13 \pm 0.19$), LOF ($0.06 \pm 0.09$), and deep autoencoder ($0.50 \pm 0.34$). The non-generalization between the unsupervised methods in a cross-evaluation fashion (i.e., inter-dataset evaluation) is similar to the findings of \cite{non_generalize_unsup}. 

We confirmed that the deep autoencoder in a federated learning setup does not guarantee a good generalization performance. The reported F1-score supports this conclusion on the 10\textsuperscript{th} round for both cases, $0.47$ considering both benign and attack thresholds, and $0.51$ for only the benign threshold. Despite \cite{TIAN2021102344} not focusing on performance between heterogeneous networks, the deep autoencoder alone does not show generalization. These obtained metrics are close to EFC's average cross-silo performance but do not outperform EFC.


As reported in Table~\ref{tab:f1-results}, our proposal, stacking of the autoencoder and EFC, was tested in two possible usages: using two thresholds for benign and attack samples (best performer, $0.84$) and only one threshold for just benign samples ($0.77$).  In both cases, the proposed method outperforms all unsupervised methods under consideration.

Based on our proposal metrics, there is a trade-off between using only the benign threshold or both. Despite a lower performance for the only benign in contrast to using both benign and attack thresholds, the only benign approach gives an advantage because we can deploy the solution relying only on benign network flow samples, which is more straightforward from an operational perspective. The EFC can also be trained only on benign samples, feasible on our stacking architecture.

\subsection{Error Analysis}
In this section, we further analyze the performance of each silo as part of the federated learning setup. We dig further into the silos performance on the 10\textsuperscript{th} round using the FedAvg strategy. The performance summary is presented in Table~\ref{tab:metrics_per_silo}. From Table~\ref{tab:metrics_per_silo}, we can confirm by the F1-score that TON-IoT (sampled only) and UNSW-NB15 are those silos that negatively impact the overall performance.

The error analysis is a systematic approach that checks our setup's mistakes on the test set. From the UNSW-NB15, these misclassified examples consist of 99\% from the Benign class out of 5.949 mislabeled samples (the test set for the sampled case is composed of 200.000 samples). Reviewing these mislabeled samples results in the majority of cases with source IP address as $175.45.176.0/24$, which according to the testbed architecture~\cite{unsw-nb15} were expected to provide only attack samples. This misbehavior must be further investigated.

For the TON-IoT sampled case, we had  52.031 misclassified examples out of 200.000. An error analysis in these cases represents that $71\%$ are scanning attacks, and $24\%$ are benign cases. From the scanning class, the misclassified samples on its majority are from IP sources $192.168.1.{30, 31, 32}$ targeting IoT/IIoT devices in the subnet $192.168.1.0/24$. According to the testbed~\cite{toniot_network_features} performs the scanning attacks with both Nmap and Nessus. Our analysis raises the question that vulnerability scanners (e.g., Nessus) can be mistakenly understood as benign traffic and must be further investigated on flow-based security mechanisms, or separated from port-scan attacks, in this case, those performed with the Nmap.        

\begin{table*}[htb]
\centering
\caption{Learning metrics for each silo during evaluation step on the 10\textsuperscript{th} round of federated learning training -- Results obtained with FedAvg strategy on sampled and reduced silos configuration (separated by ``/'') with both benign and attack threshold.}
\label{tab:metrics_per_silo}
\resizebox{\textwidth}{!}{%
\begin{tabular}{@{}rccccccc@{}}
\toprule
\textbf{Silo}   & \textbf{Accuracy} & \textbf{Recall} & \textbf{Precision} & \textbf{F1-Score} 
& \textbf{Miss-rate} & \textbf{Fallout} & \textbf{AUC} \\ \midrule
Bot-IoT         &  0.93 / 0.90                &  0.93 / 0.90              &   0.99 / 0.99                 &  0.96 / 0.95                 
&       0.07 / 0.1            &    0.02 / 0.01              &     0.95 / 0.94         \\
TON-IoT         &    0.74 / 0.85              &     0.69 / 0.85           &     0.87 / 0.95              &   0.77 / 0.90               
&     0.31 / 0.14              &    0.18 / 0.14             &     0.76 / 0.86        \\
UNSW-NB15       &      0.97 / 0.97            &     0.99 / 1.0           &   0.57 / 0.50                &    0.73 / 0.66              
&         0.0002 / 0.0          &       0.03 / 0.14          &       0.98 / 0.99      \\
CSE-CIC-IDS-2018 &      0.98 / 0.98            &    0.88 / 0.88            &    0.91 / 0.92              &   0.90 / 0.90               
&     0.11 / 0.12              &    0.01 / 0.01             &      0.94 / 0.93       \\ \bottomrule
\end{tabular}%
}
\end{table*}

This error analysis could derive further actions such as the detailed investigation of the original UNSW-NB15 dataset and the use of vulnerability scanners in the same category as port scanners on the TON-IoT dataset. Thus, working on cases of UNSW-NB15 and TON-IoT (e.g., understanding the network or the specific composition of the datasets) would achieve the most significant improvement of our method. Similar discussion about troubleshooting NIDS dataset is presented by~\citep{troubleshoot_cicids}. Although, our proposed method is not expected to be tied with the specific datasets chosen in this paper. So, fixing the mislabeled samples on the dataset is out of scope. Additionally, this post-mortem analysis is insightful for interpretability, but addressing the changes in our current data would introduce data snooping to the overall analysis. However, the insights possible with this error analysis are recommended to be part of the deployment, composing the overall operations of a federated learning-based intrusion detection system.

\section{Conclusions}\label{conclusion}
In this work, we present unsupervised federated learning (FL)-based network intrusion detection system (NIDS). It can achieve generalization between heterogeneous networks in a stacking setup with a descriptive algorithm, the Energy Flow Classifier (EFC). The heterogeneous networks are represented in our evaluation as four recent NIDS datasets from different network contexts.
This generalization advances the research of NIDS capable of being deployed in diverse environments. Additionally, our unsupervised approach can be deployed in network environments with only benign network flow-based samples.

We contribute by bonding the field of FL-based NIDS and the generalization of machine learning-based NIDS. With the evidence of FL as a promising approach for generalization between diverse networks. Also, with the use of FL, we provide an efficient distributed NIDS, once no data is required to be moved to a central server, and provide privacy for the system participants.

In future works, we plan to investigate the proposed method in a cross-device environment, considering the deployment of a distributed NIDS in conjunction with the concept of zero-trust architecture. Additionally, using FL introduces new security challenges related to adversarial machine learning. Thus, we consider the robustness evaluation of our proposal for adversarial attacks as future works for a dependable distributed NIDS solution.



\bibliographystyle{elsarticle-num} 
\bibliography{bibfile}





\end{document}